\documentclass[float, aps, pre, twocolumn, superscriptaddress,groupedaddress, letter, longbibliography]{revtex4-1}  \usepackage{graphicx}  
\DeclareGraphicsExtensions{.png,.pdf,.eps}

\usepackage{dcolumn}   \usepackage{bm}        \usepackage{amsmath}   \usepackage{comment}   
\usepackage{xcolor}
\usepackage{rotating}

\newcommand\crule[1][black]{\textcolor{#1}{\rule{0.5em}{0.5em}}}

\newcommand{\bu}{{\bf u}}

\newcommand{\vx}{\mathbf{x}}
\newcommand{\sfp}{p}

\newcommand{\sfPz}{p_0}

\newcommand{\vfu}{\mathbf{u}}
\newcommand{\vff}{\mathbf{f}}
\newcommand{\vfuhat}{\hat{\mathbf{u}}}
\newcommand{\vffhat}{\hat{\mathbf{f}}}
\newcommand{\vfU}{\mathbf{U}}
\newcommand{\vfUz}{\mathbf{U}_0}
\newcommand{\vffz}{\mathbf{f}_0}

\newcommand{\vdiv}{\nabla \cdot}

\newcommand{\vgrad}{\nabla}

\newcommand{\dd}[2]{\frac{\partial{#1}}{\partial{#2}}}

\newcommand{\resolvent}{\mathcal{H}}

\begin{document}

\title{Low-dimensional representations of exact coherent states of the Navier-Stokes equations from the resolvent model of wall turbulence}

\author{Ati S. Sharma}
\email{a.sharma@soton.ac.uk}
\affiliation{Unversity of Southampton, UK}

\author{Rashad Moarref}
\email{rashad@caltech.edu}
\affiliation{California Institute of Technology, USA}

\author{Jae Sung Park}
\email{park329@wisc.edu}
\affiliation{Unversity of Wisconsin-Madison, USA}

\author{Michael D. Graham}
\email{graham@engr.wisc.edu}
\affiliation{Unversity of Wisconsin-Madison, USA}

\author{Beverley J. McKeon}
\email{mckeon@caltech.edu}
\affiliation{California Institute of Technology, USA}

\author{Ashley P. Willis}
\email{a.p.willis@sheffield.ac.uk}
\affiliation{Unversity of Sheffield, UK}

\date{\today}

\begin{abstract}
                    
    We report that many exact invariant solutions of the Navier-Stokes equations for both pipe and channel flows are well represented by just few modes of the model of McKeon \& Sharma {J.~Fl.~Mech.} \textbf{658}, 356 (2010). This model provides modes that act as a basis to decompose the velocity field, ordered by their amplitude of response to forcing arising from the interaction between scales. The model was originally derived from the Navier-Stokes equations to represent turbulent flows and has been used to explain coherent structure and to predict turbulent statistics.
    This establishes a surprising new link between the two distinct approaches to understanding turbulence.
    
\end{abstract}

\maketitle

The problem of finding simple predictive descriptions of turbulence has endured since at least the time of Reynolds. Recently, two viewpoints have emerged that explain structure in turbulence in quite different ways: firstly, in terms of invariant solutions of the Navier-Stokes equations, which has been used to explain the transition to turbulence; secondly, in terms of selective amplification or filtering of a superposition of travelling waves. In this paper we show that the latter approach efficiently captures the structure of these invariant solutions, providing a new and surprising link between the two distinct approaches and supporting the idea that these invariant solutions share the same dominant mechanisms as flows in the turbulent regime.

The first viewpoint comes from treating the infinite-dimensional Navier-Stokes equations that govern turbulence as a nonlinear dynamical system. The programme of work arising from this viewpoint has centred on finding invariant solutions of the Navier-Stokes equations that appear constant in a co-moving frame of reference \cite{Waleffe1998, FaisstEckhardt2003, WedinKerswell2004}, and on finding periodic orbits \cite{KawaharaKida2001, Viswanath2007, CvitanovicGibson2010}. It is hoped that such invariant solutions may eventually be used in a weighted expansion to compactly describe turbulent flows \cite{Kerswell2005}. 

 These invariant solutions arise in pairs at finite amplitude via a saddle-node bifurcation at a particular Reynolds number.  The so-called lower branch (L) solution of each pair denotes a state with lower drag than its corresponding upper branch (U) solution.
These solutions are thought to underlie the structure of turbulence by concentrating state space trajectories in their vicinity. Although the dynamical systems description has been most successful at describing
transitional flows, it has been argued that such solutions are relevant to turbulent flow \cite{Waleffe1997, WangGibson2007} and recent experimental evidence supports the view that these solutions continue to be important in turbulent flows and are ultimately responsible for turbulent statistics \cite{DennisSogaro2014}.

The second viewpoint is the model of McKeon \& Sharma which arose from systems and control theory \cite{McKeonSharma2010, McKeonSharma2013}. This approach treats turbulence as a superposition of travelling waves, which are attenuated or amplified according to their interaction with the mean flow, and excited by other travelling waves.
 In this model, the structure and robustness of turbulence comes from the interplay between this linear amplification and an energy-conserving nonlinear feedback mechanism. The resolvent formulation generates an ordered set of basis functions by choosing the velocity fields arising from the most amplified forcing, the next most, and so forth. The model has been used to make predictions about the spatial organisation of turbulent velocity fluctuations \cite{SharmaMcKeon2013} and turbulent fluctuation energy spectra \cite{MoarrefSharma2013, MoarrefJovanovic2014}. The resulting modes are travelling waves with phase and amplitude that varies spatially.
Unlike approaches such as Dynamic Mode Decomposition \cite{Schmid:2010}, Proper Orthogonal Decomposition \citep{Berkooz.Holmes.Lumley:1993}, or wavelets \citep{Farge:1992}, the model is derived from the equations rather than from an existing data set.
Notably, this viewpoint is entirely in the frequency-domain; kinematic descriptions are abandoned in favour of a system-level selection of travelling waves.
 The origin of these basis functions has a clear physical interpretation. The mechanisms are high amplification at the critical layer, where the phase velocity equals the flow velocity; the lift-up mechanism, where the flow velocity fluctuations extract energy using the shear in the mean flow; and high amplification for modes with long streamwise wavelength. 

The presence of only one phase velocity in the invariant solutions used here greatly simplifies the problem of comparison to the resolvent formulation, in contrast to difficulties encountered in the turbulent case \cite{GomezBlackburn2014}. Thus, the frequency-domain view of turbulence as a superposition of interacting travelling waves is well suited to the analysis of exact solutions.

Both the control theory viewpoint and the nonlinear dynamics invariant solutions
 viewpoint bring different and important insights, so unifying these distinct approaches would be an important advance in our understanding of turbulence.
In this Rapid Communication, we show that the exact invariant solutions often are well represented by a relatively small number of model modes. 
This shows that the same mechanisms are dominant in the invariant solutions as in the model, and therefore as in turbulent flows.

In the following, we project exact invariant solutions in pipe and channel flow onto basis functions (modes) generated by the model from the mean velocity profile of the solutions.
We use the notation $U_B$ for the bulk velocity, $R$ for the pipe radius, $h$ for the channel half-height, $u_\tau$ for the friction velocity and $\nu$ for the kinematic viscosity. 

The pipe solutions, presented first, were generated by continuation using the pseudo-arclength method to $Re_B=2 U_B R / \nu = 5300$ ($Re_\tau=u_\tau h / \nu =106-214$)
from the solutions of \cite{PringleDuguet2009} using 
\emph{openpipeflow.org}. The wall-normal resolution was 60 points.
These solutions are classified into N-class and S-class.
The N-class solutions have mirror, shift-and-reflect and rotational symmetries, with wavy fast streaks and slow streaks arranged to interact with quasi-streamwise vortices. The S-class have only shift-and-reflect symmetry, but are otherwise similar in structure.
Six S-class solutions and ten N-class solutions were used, of which four were upper branch and the rest lower branch. The N-class upper branch solutions have a friction factor close to that of turbulent flow, whereas the others are close to laminar flow.

The channel solutions are from families P1, P3 and P4 of \cite{pargra15} and were generated using the code \emph{channelflow} \cite{GibsonHalcrow2007}. The wall-normal resolution was 81 points. The P1 (at $Re_\tau = 75$) and P3 (at $Re_\tau=85$)  families are active in the core of the channel, and approach laminar as Reynolds number increases. There is as yet no widely accepted theory for the mechanism that drives these solutions.
The P4 solutions (at $Re_\tau=85$) are highly nonlinear with fluctuations localised near the critical layer. Their sustaining mechanism is well understood \cite{WangGibson2007, HallSherwin2010}. The critical layer for these solutions varies spatially.
The P1 and P3 lower branch solutions have been continued to higher Reynolds number by the pseudo-arclength method. For these solutions, the importance of the critical layer mechanism becomes clearer at much higher Reynolds number \cite{Viswanath2009, HallSherwin2010, DeguchiHall2014}.

The systems model from which the basis functions derive is formulated from the Navier-Stokes equations as follows. In the following, the three-component velocity field is denoted by $\vfU(\vx,t)$ and the long time-averaged velocity field is denoted by $\vfUz(\vx)$, with $\vx$ being a point in the flow interior and $t$ being time. The mean velocity $\vfUz$ and associated pressure $\sfPz$ are assumed known \emph{a priori}. The fluctuations are then $\vfu=\vfU-\vfUz$. The Navier-Stokes equations can be put in the form
\begin{align}
    \dd{ \vfu }{t} &= -\vgrad \sfp  - \vfUz\cdot\vgrad\vfu - \vfu\cdot\vgrad\vfUz + Re^{-1}\vgrad^2 \vfu + \vff\\
    \vffz &= \vfUz\cdot\vgrad\vfUz + \vgrad \sfPz - Re^{-1}\vgrad^2\vfUz  \\
    \vff &= -\vfu\cdot\vgrad\vfu\\
    0 &= \vdiv \vfu = \vdiv \vfUz.
\end{align}

The model formulation proceeds by considering a superposition of fluctuations in an infinite pipe or channel, of purely harmonic form at temporal frequency $\omega$, streamwise wavenumber $\alpha$, and azimuthal (spanwise) wavenumber $\beta$, allowing the first equation (linear in the fluctuations) to be considered as harmonic disturbances forced by the interaction between other harmonic disturbances. The phase velocity is then $c=\omega/\alpha$.
The equation for the fluctuations is then of the form
\begin{equation}
    \vfuhat(y; \alpha, \beta, c)=\resolvent_{\alpha,\beta,c} \ \vffhat(y; \alpha, \beta, c)
\end{equation}
where $y$ is the wall-normal distance and the $\hat{~}$ notation indicates the appropriate complex Fourier coefficient.
The object of the analysis is the linear operator $\resolvent_{\alpha,\beta,c} $, which is known as the resolvent operator. The analysis then considers the singular value decomposition of $\resolvent$,
\begin{equation}
    \resolvent_{\alpha,\beta,c} =\sum_m \psi_m(y;\alpha,\beta,c) \ \sigma_m(\alpha,\beta,c) \ \phi^*_m(y;\alpha,\beta,c)
\end{equation}
By definition, the left and right singular vectors and the singular values obey the orthogonality and ordering conditions, $\left(\phi_m(y;\alpha,\beta,c), \phi_{m'}(y;\alpha,\beta,c) \right)_y=\delta_{m,m'}$, $\left(\psi_m(y;\alpha,\beta,c),\  \psi_{m'}(y;\alpha,\beta,c) \right)_y=\delta_{m,m'}$, $\sigma_m \geq \sigma_{m+1}$.

The singular values $\sigma_m$ are the amplification factors from $\vffhat$ to $\vfuhat$ and the left singular vectors $\psi_m$ are the basis functions (response modes) which represent the velocity field. The singular values each represent the gain from forcing with the associated right singular vector. This gain is assumed to rank the importance of a mode pair in a flow, and thus induces a natural ordering of the modes.

Given the particular ${\bf U}_0$ for each invariant solution this therefore results in response modes $\psi_m$ particular to each solution onto which it may be projected, with $\sigma_m$ indicative of the importance of each mode. Only modes with the appropriate phase velocity need to be considered. To the extent that the modes and singular values correctly capture the relevant physics of the solution, only a small number of modes will be needed.

\begin{figure}
    \begin{minipage}[t]{0.45\columnwidth}
        \begin{tabular}[b]{c|ccc}
            class & ~$\alpha$  &  $Re_\tau$ & $c/2U_B$    \\ \hline
            S1  & 0.78  & 106  & 0.81      \\
            S1  & 0.34  & 107  & 0.78      \\
            S2b & 1.24  & 107  & 0.76      \\
            S2b & 0.95  & 108  & 0.76     \\
            S2b & 0.39  & 110  & 0.71     \\
            S2b & 0.24  & 122  & 0.62     \\
            N2L & 1.73  & 109  & 0.83     \\
            N2L & 1.24  & 109  & 0.82     \\
            N2L & 0.55  & 111  & 0.78     \\
            N2L & 0.15  & 117  & 0.70     \\
            N3L & 1.25  & 111  & 0.74     \\
            N4L & 1.70  & 114 & 0.69      \\
            N2U & 1.25  & 151 & 0.66      \\
            N2U & 1.01  & 156 & 0.66      \\
            N2U & 0.80  & 177 & 0.64      \\
            N4U & 1.70  & 214 & 0.55      \\
            \hline
            class   & $Re_B$ & $Re_\tau$ & $c/2U_B$   \\ \hline
            P1L    & 3200 & 75 & 0.58     \\
            P3L    & 4533 & 85 & 0.52     \\
            P4L    & 3677 & 85 & 0.58     \\
            P1U    & 2267 & 75 & 0.49     \\
            P3U    & 1733 & 85 & 0.35     \\
            P4U    & 2200 & 85 & 0.55     \\
            \hline
        \end{tabular}
    \end{minipage}
    \begin{minipage}[t]{0.43\columnwidth}
        \includegraphics[width=\columnwidth]{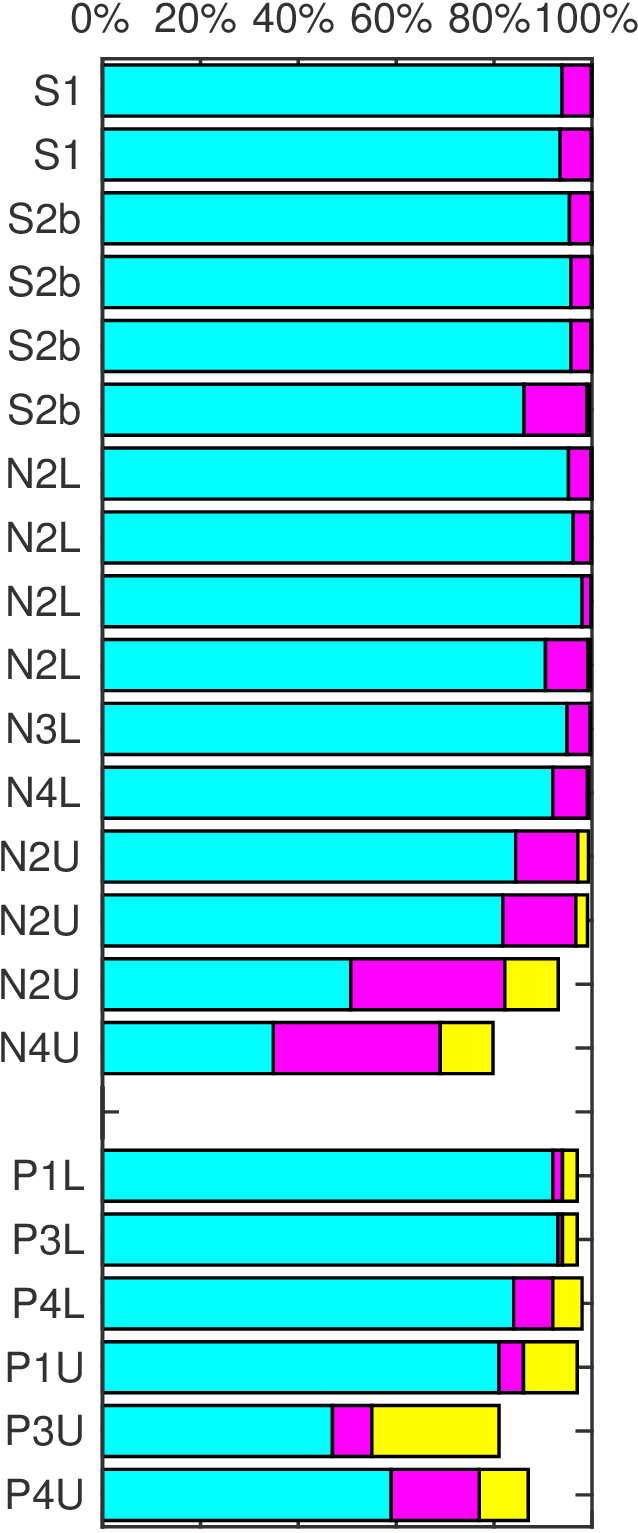}
    \end{minipage}
    \caption{
        (Color online) The upper set refer to the pipe solutions (all at $Re_B=5300$), the lower set to the channel solutions (at a range of $Re_B$).
  \emph{Left:} All invariant solutions considered in this study, covering a range
  of solution classes, $Re_B$, $Re_\tau$ and wavespeeds $c$. \emph{Right, upper set:} Fraction of energy captured by a projection of $m=1,5,10$ model modes per Fourier mode (\crule[cyan],\crule[magenta],\crule[yellow]; pipe solutions). \emph{Right, lower set:} $m=1,2,5$ model mode pairs per Fourier mode (\crule[cyan],\crule[magenta],\crule[yellow]; channel solutions).
  }
    \label{proj-E-all}
\end{figure}

\begin{figure}
    \begin{minipage}[t]{0.45\columnwidth}
        \begin{tabular}[b]{c|cc}
            class & $Re_\tau$  & $\varepsilon$  \\ \hline
            S1  & 106 & 1.07\\
            S1  & 107 & 1.09\\
            S2b & 107 & 1.09\\
            S2b & 108 & 1.09\\
            S2b & 110 & 1.15\\
            S2b & 122 & 1.41\\
            N2L & 109 & 1.11\\
            N2L & 109 & 1.12\\
            N2L & 111 & 1.16\\
            N2L & 117 & 1.30\\
            N3L & 111 & 1.16\\
            N4L & 114 & 1.25\\
            N2U & 151 & 2.17\\
            N2U & 156 & 2.31\\
            N2U & 177 & 3.02\\
            N4U & 214 & 4.49\\
            \hline
            class   & $Re_\tau$ & $\varepsilon$ \\ \hline
            P1L    & 75 & 1.19 \\
            P3L    & 85 & 1.08 \\
            P4L    & 85 & 1.32 \\
            P1U    & 75 & 1.62 \\
            P3U    & 85 & 2.77 \\
            P4U    & 85 & 2.18 \\
            \hline
        \end{tabular}
    \end{minipage}
    \begin{minipage}[t]{0.43\columnwidth}
        \includegraphics[width=\columnwidth]{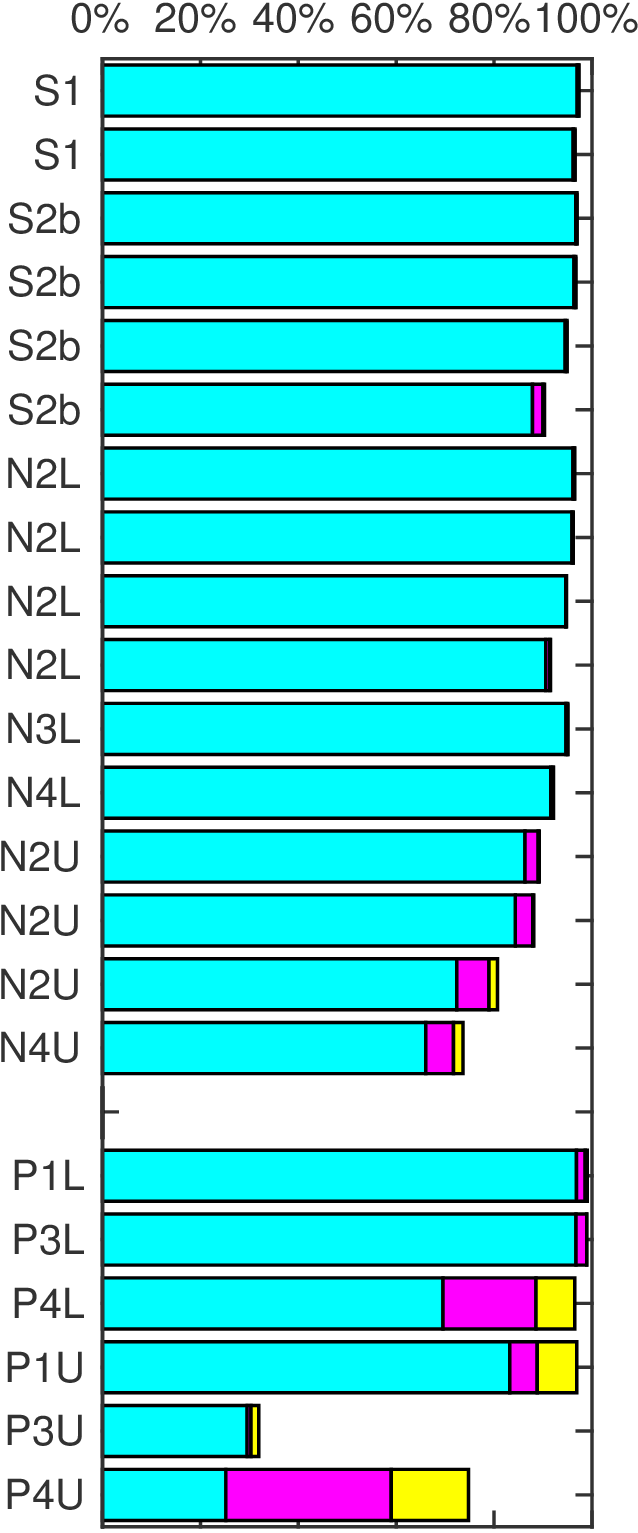}
            \end{minipage}
    \caption{
        (Color online) The upper set refer to the pipe solutions (all at $Re_B=5300$), the lower set to the channel solutions (at a range of $Re_B$).
  \emph{Left:} All invariant solutions considered in this study. The quantity $\varepsilon$ is the internal dissipation of fluctuations relative to the laminar flow with the same bulk velocity. \emph{Right, upper set:} Fraction of total internal dissipation $\varepsilon$ due to fluctuations captured by a projection of $m=1,5,10$ model modes per Fourier mode (\crule[cyan],\crule[magenta],\crule[yellow]; pipe solutions). \emph{Right, lower set:} $m=1,2,5$ model mode pairs per Fourier mode (\crule[cyan],\crule[magenta],\crule[yellow]; channel solutions). 
    }
    \label{proj-eps-all}
\end{figure}

\begin{figure*}
    \centering
    \includegraphics[height=0.45\columnwidth]{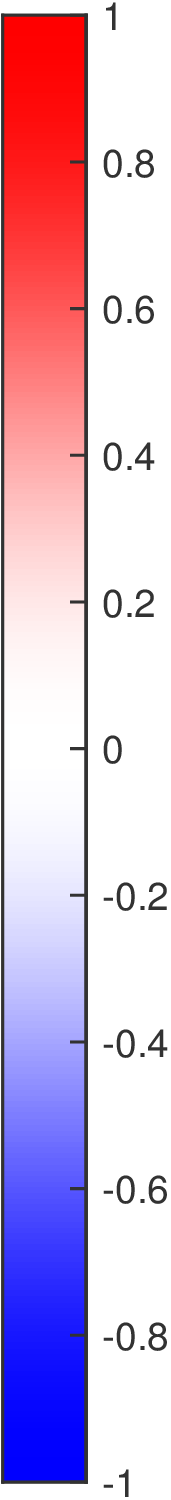}
    \includegraphics[width=0.45\columnwidth]{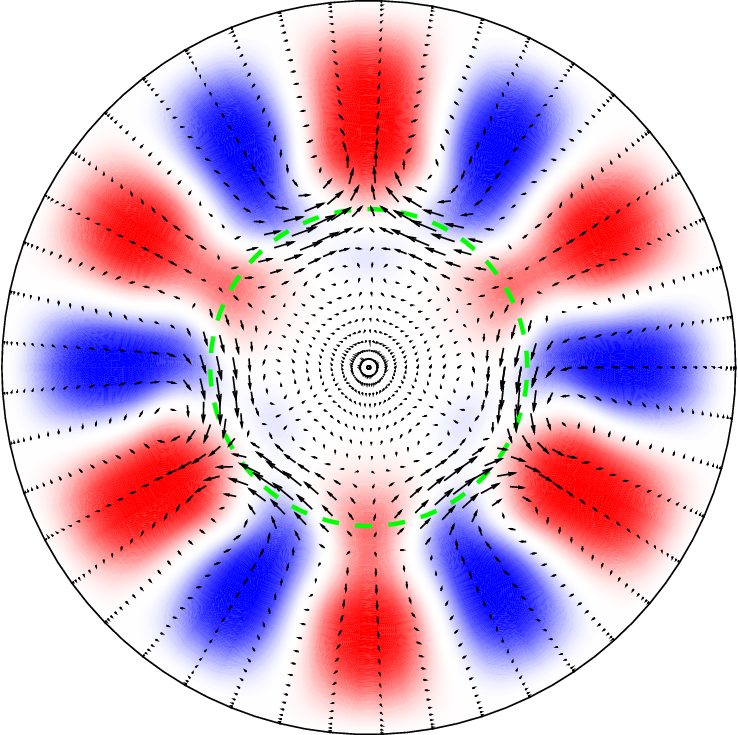}
    \includegraphics[width=0.45\columnwidth]{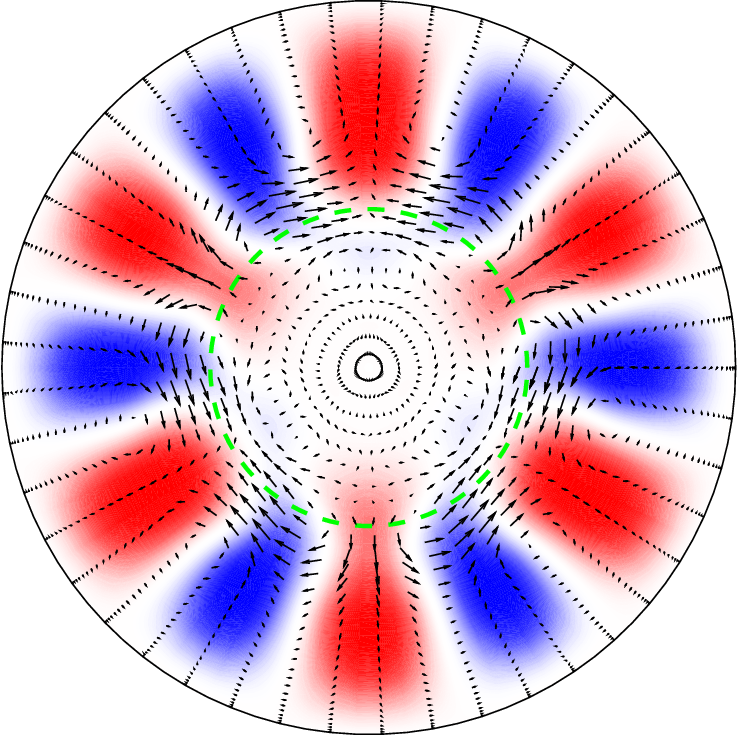}
    \includegraphics[width=0.45\columnwidth]{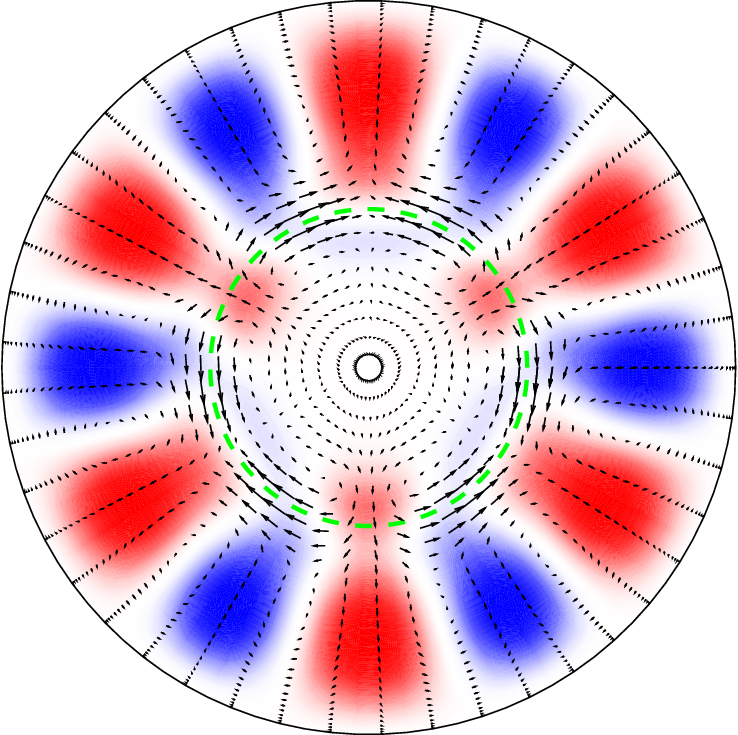}
    \includegraphics[height=0.45\columnwidth]{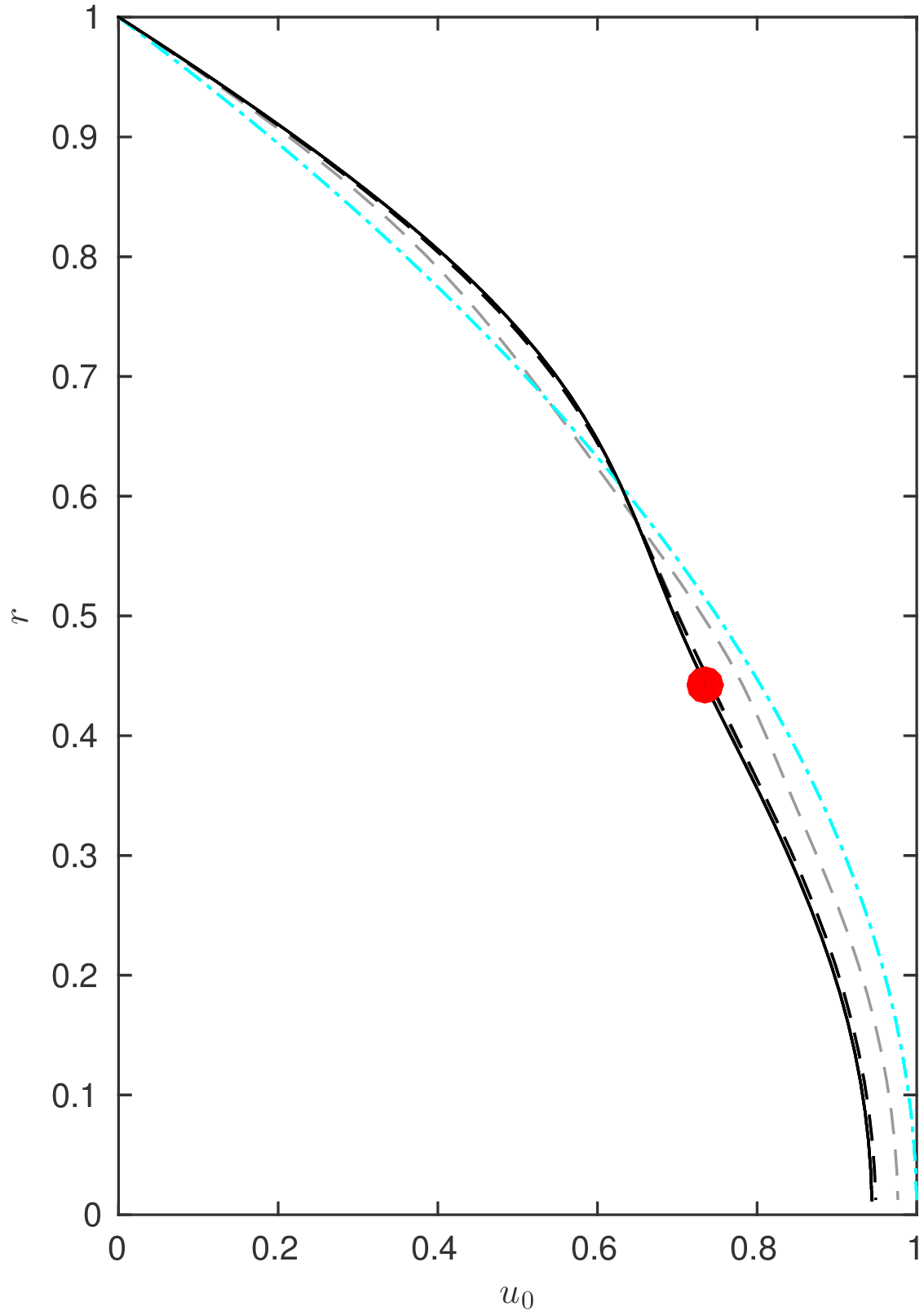}
    \caption{(Color online) N3L, lower branch solution in a pipe. From left to right: actual solution; projection onto five response modes per Fourier mode (containing 98\% of the fluctuation energy); projection onto one response mode per Fourier mode (containing 95\% of the fluctuation energy); mean velocity profile.
   The red and blue shading indicates streamwise velocity fluctuation faster and slower than the mean velocity, respectively (as a fraction of the maximum amplitude streamwise velocity). The quiver arrows indicate in-plane velocity. The wall-normal region where the phase velocity is closest to the mean velocity is indicated by a dashed green line in the pipe cross-sections and a red dot indicates the phase velocity in the mean velocity profile. The lower branch solutions such as this one are close to laminar, as seen from the mean velocity profile.
   The mean velocity profile (---) has superposed the mean flow generated by the projections with 1, 5, 10 singular values (light to dark $-~-$) and the laminar (\textcolor{cyan}{$- \cdot$}).}
    \label{pipe-proj-6507-1000}
\end{figure*}

\begin{figure*}
    \centering
    \includegraphics[width=0.45\columnwidth]{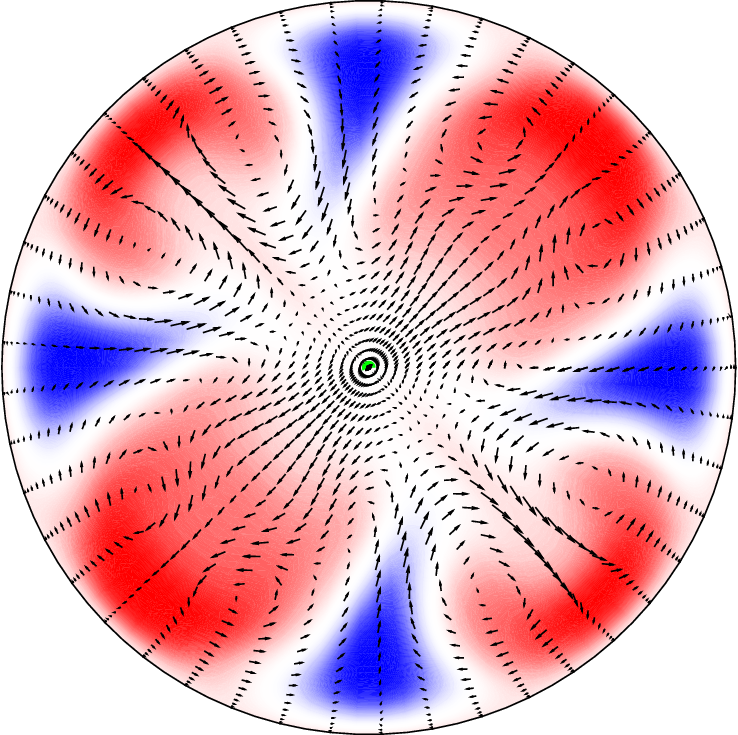}
    \includegraphics[width=0.45\columnwidth]{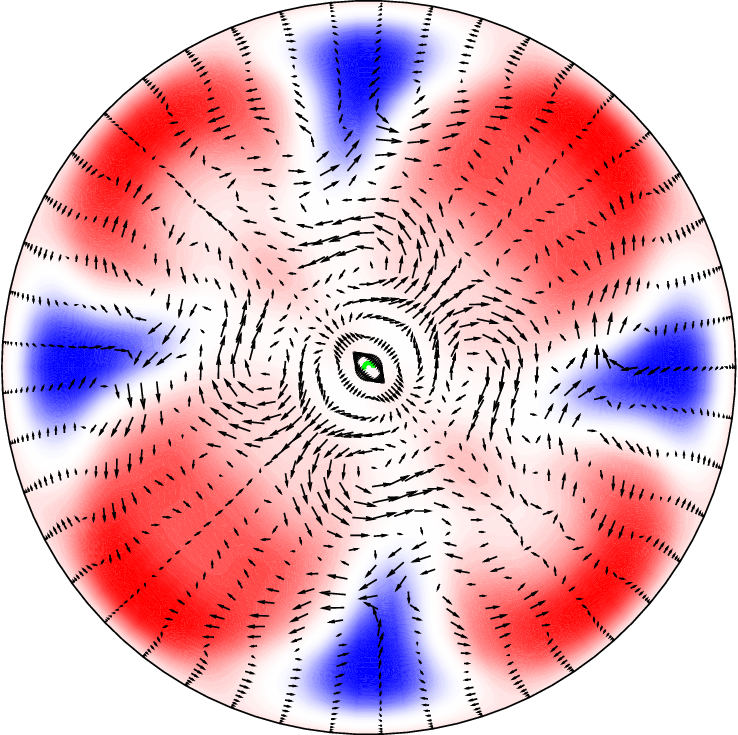}
    \includegraphics[width=0.45\columnwidth]{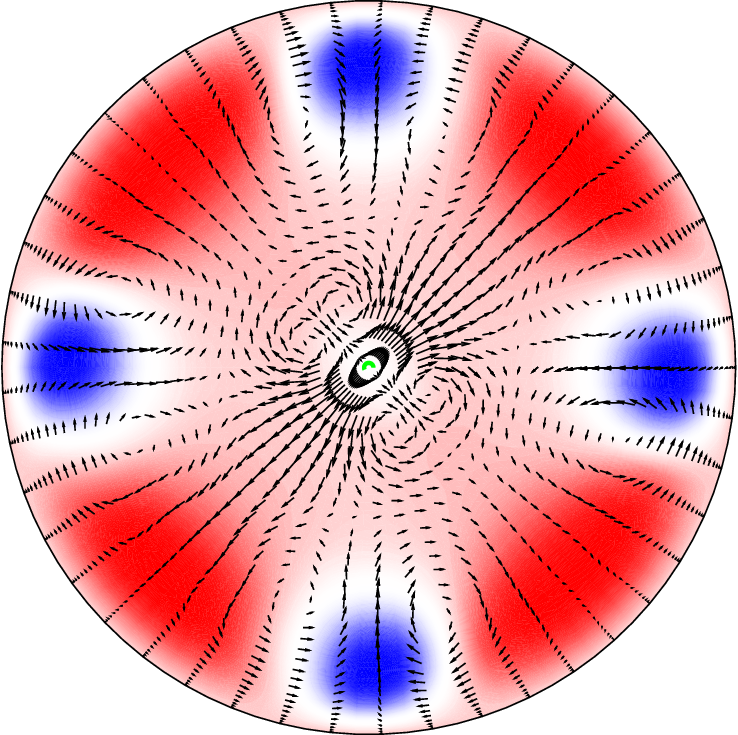}
    \includegraphics[height=0.45\columnwidth]{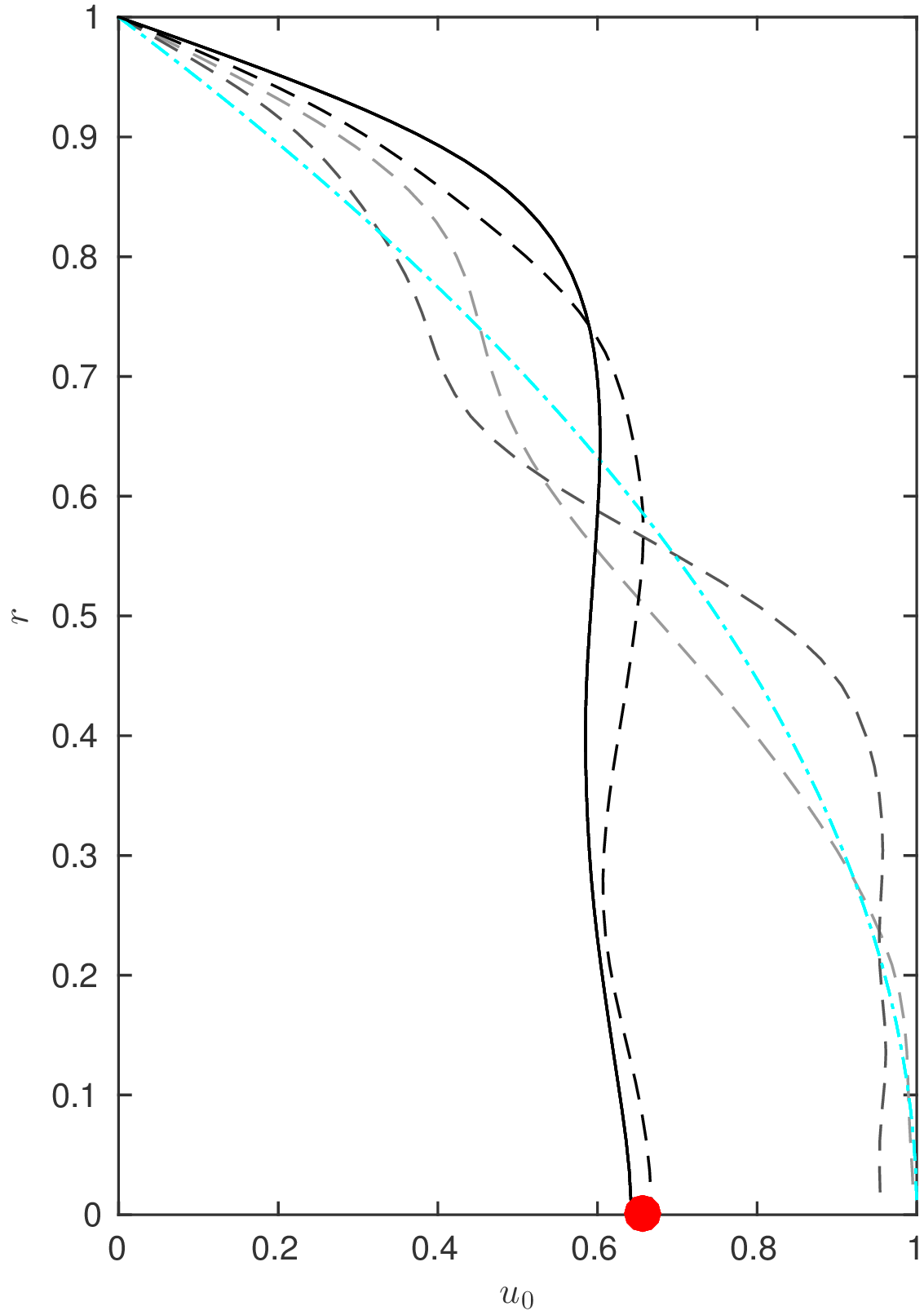}
    \caption{(Color online) N2U ($\alpha=1.25$), upper branch solution in a pipe. From left to right: actual solution; projection onto five response modes per Fourier mode (containing 96\% of the fluctuation energy), projection onto one response mode per Fourier mode (containing 84\% of the fluctuation energy); mean velocity profile. It is interesting to note that due to the flatness of the mean velocity profile, the solutions does not possess an average critical layer.}
    \label{pipe-proj-6502-0001}
\end{figure*}

A set of exact invariant solutions for channel and pipe geometries, broadly representative of all known lower and upper branch solutions with single $c$, were projected onto the modes given by the model.

\begin{figure*}
    \includegraphics[width=0.5\columnwidth]{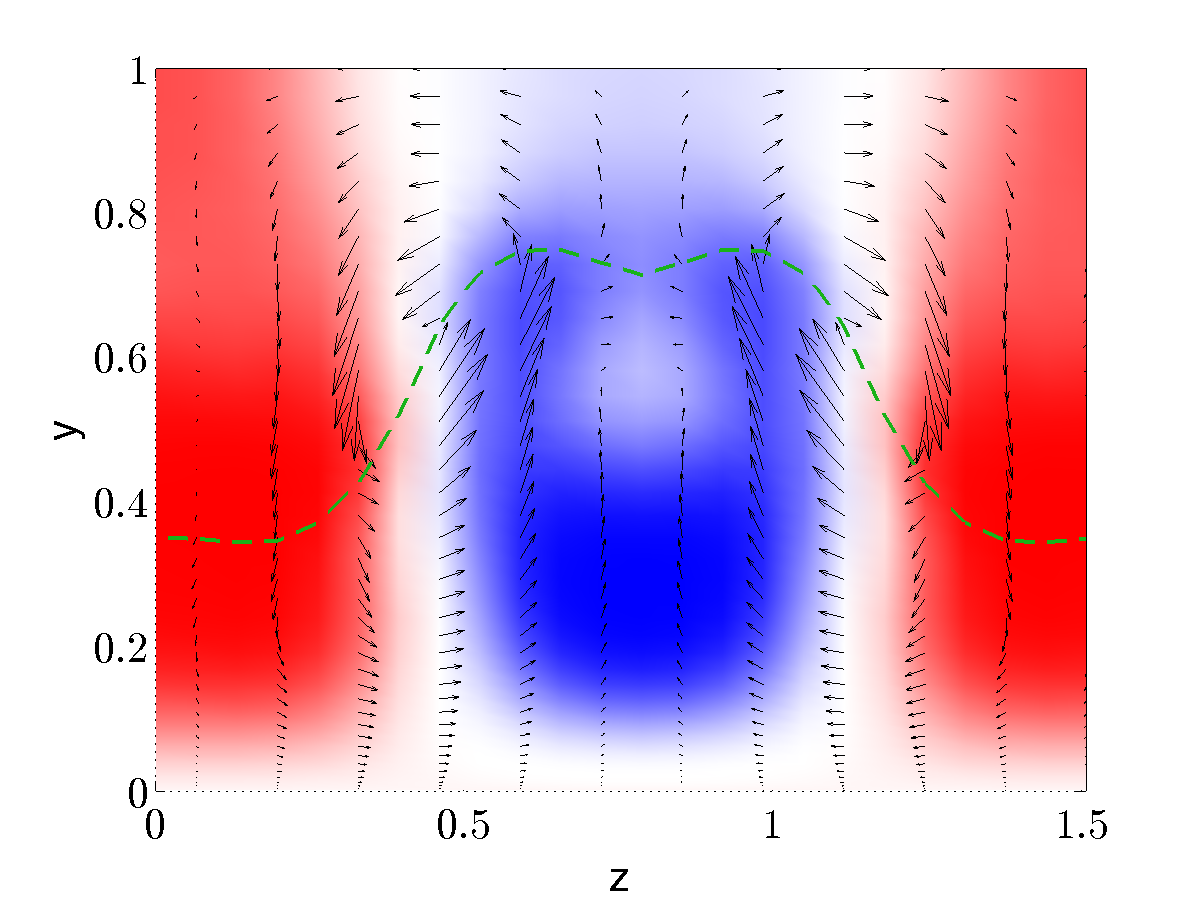}
    \includegraphics[width=0.5\columnwidth]{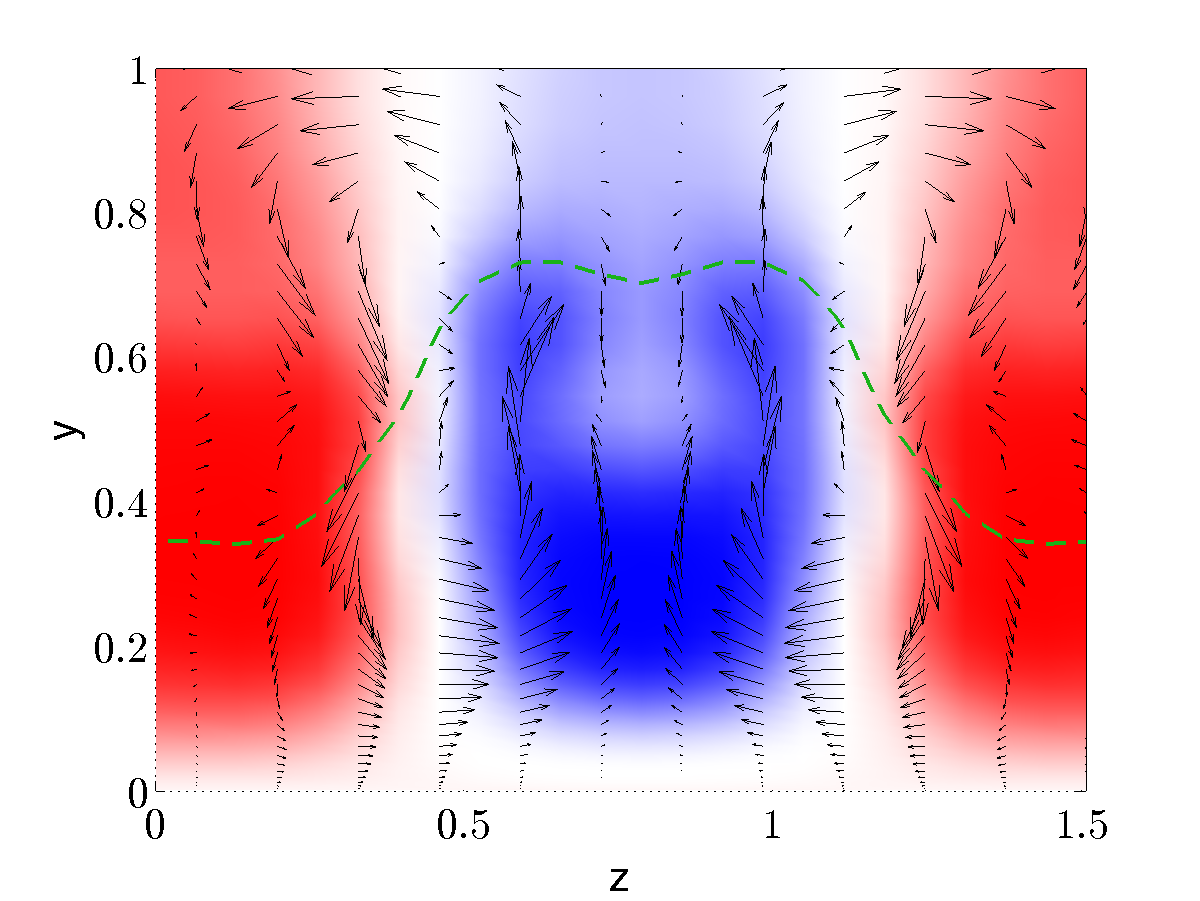}
    \includegraphics[width=0.5\columnwidth]{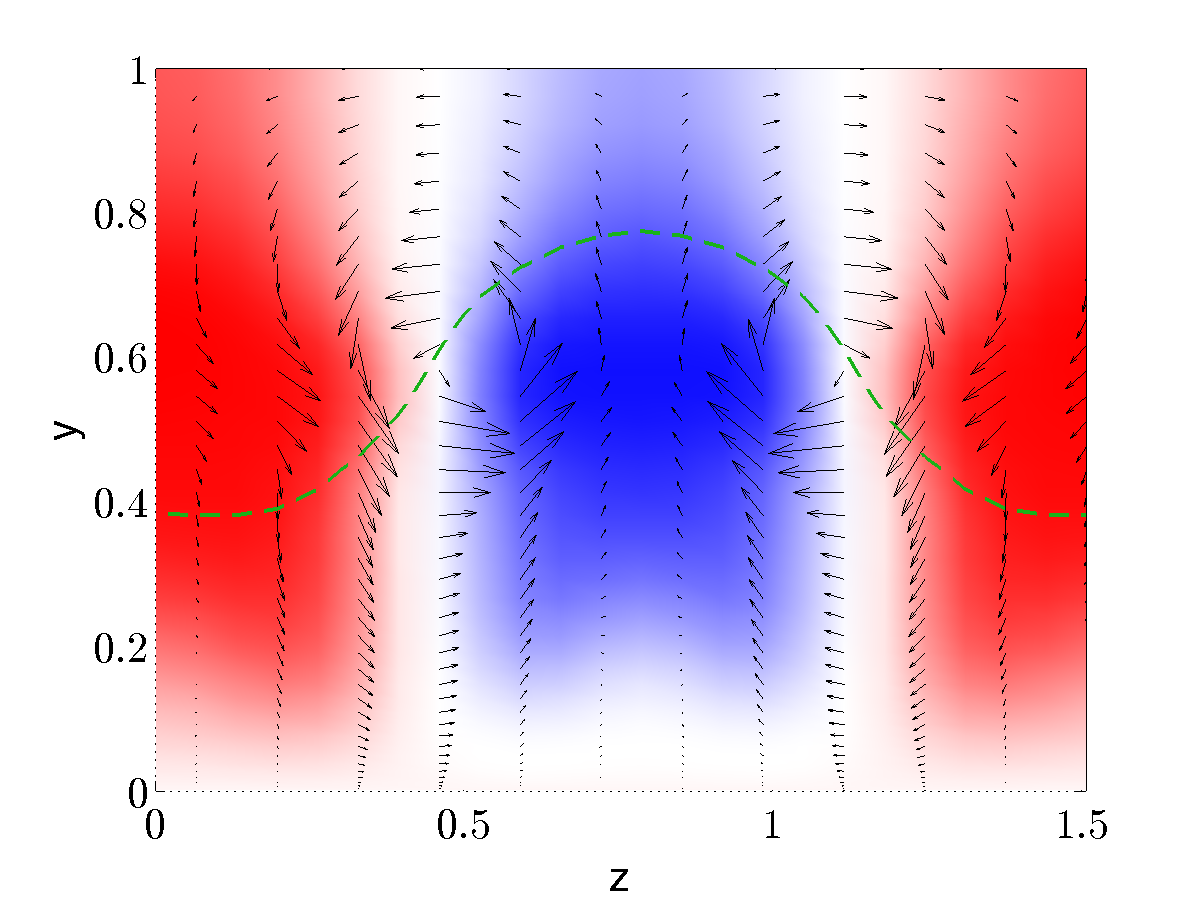}
    \includegraphics[width=0.4\columnwidth]{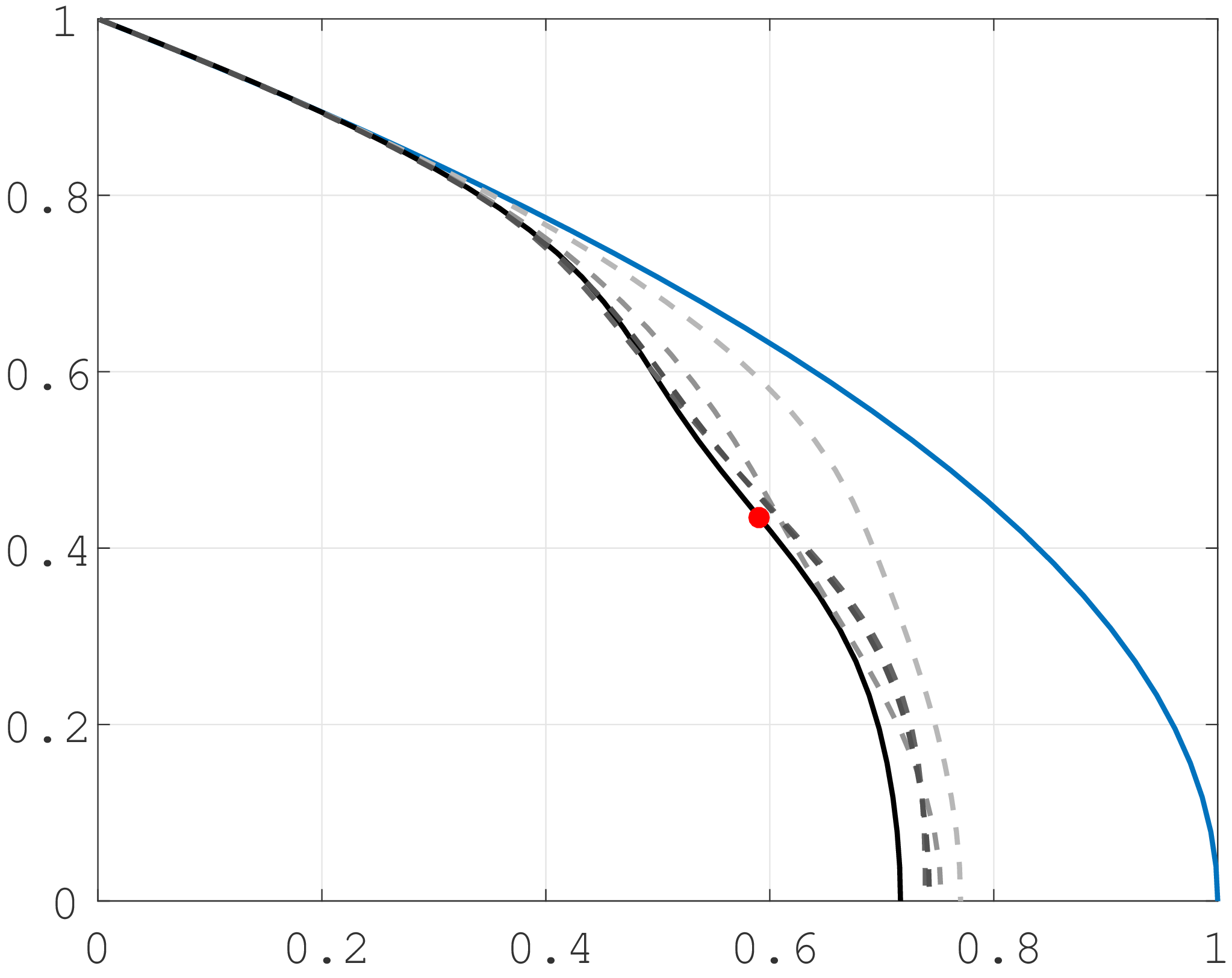}
    \caption{(Color online) P4 lower branch solution in a channel. From left to right: actual solution; projection onto five response modes pairs per Fourier mode (containing 92\% of the fluctuation energy), projection onto one response mode pair per Fourier mode (containing 84\% of the fluctuation energy); mean velocity profile.}
    \label{channel-proj-P4L}
\end{figure*}

\begin{figure*}
    \includegraphics[width=0.5\columnwidth]{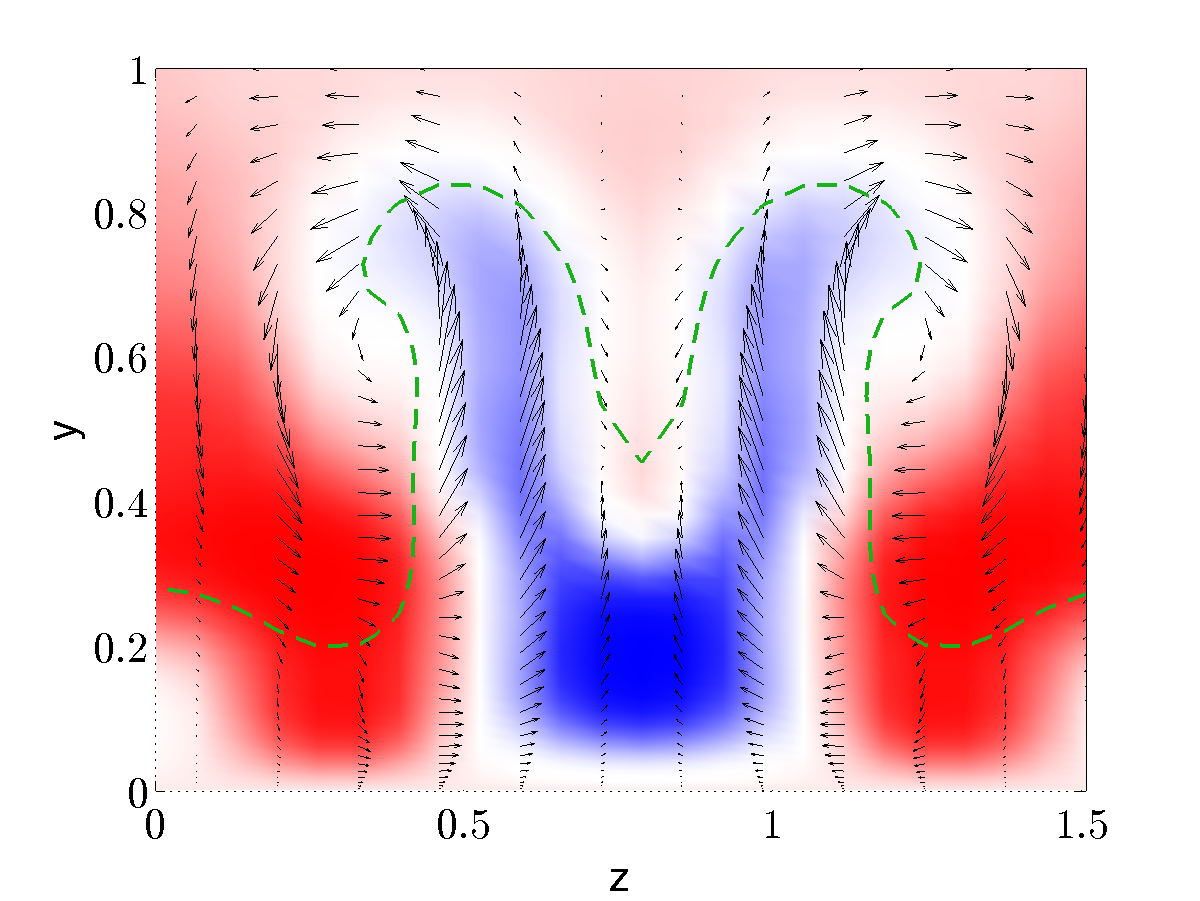}
    \includegraphics[width=0.5\columnwidth]{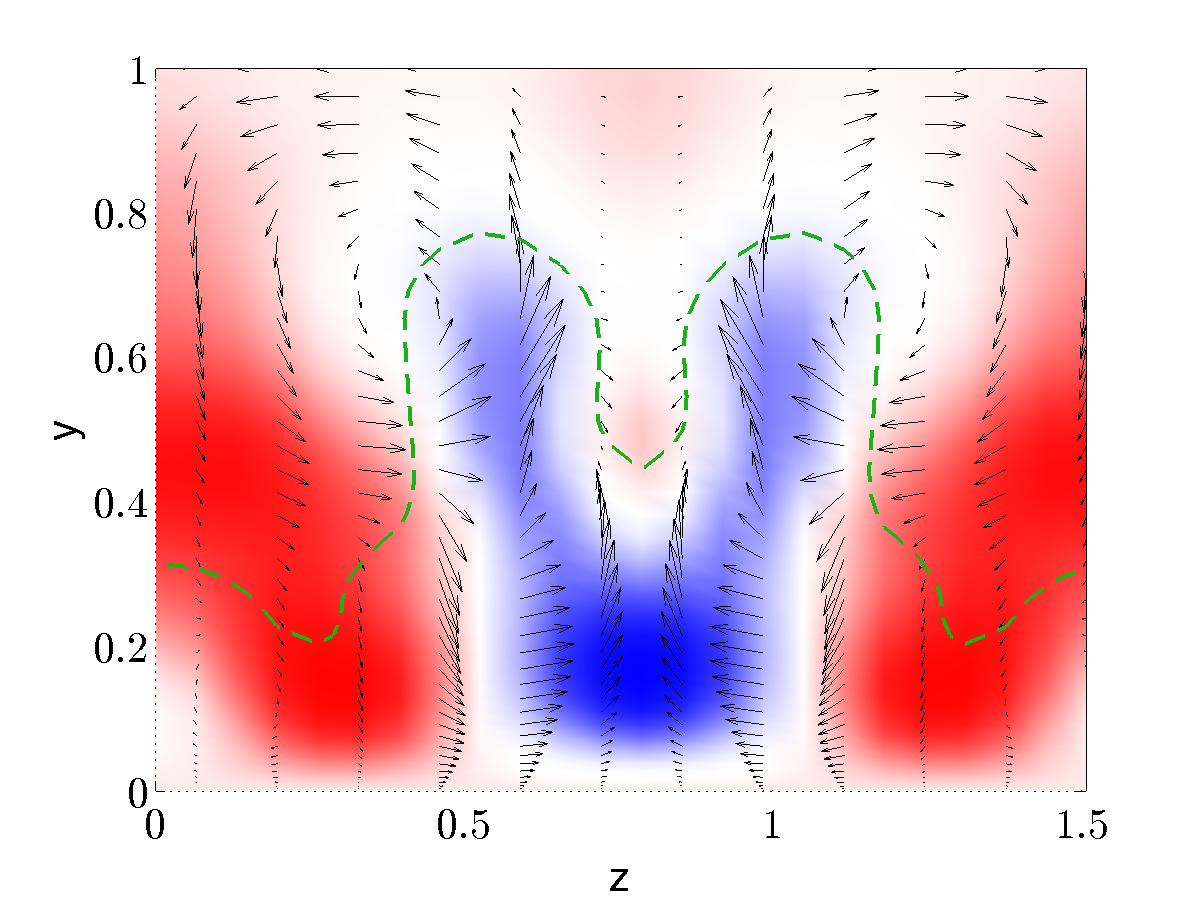}
    \includegraphics[width=0.5\columnwidth]{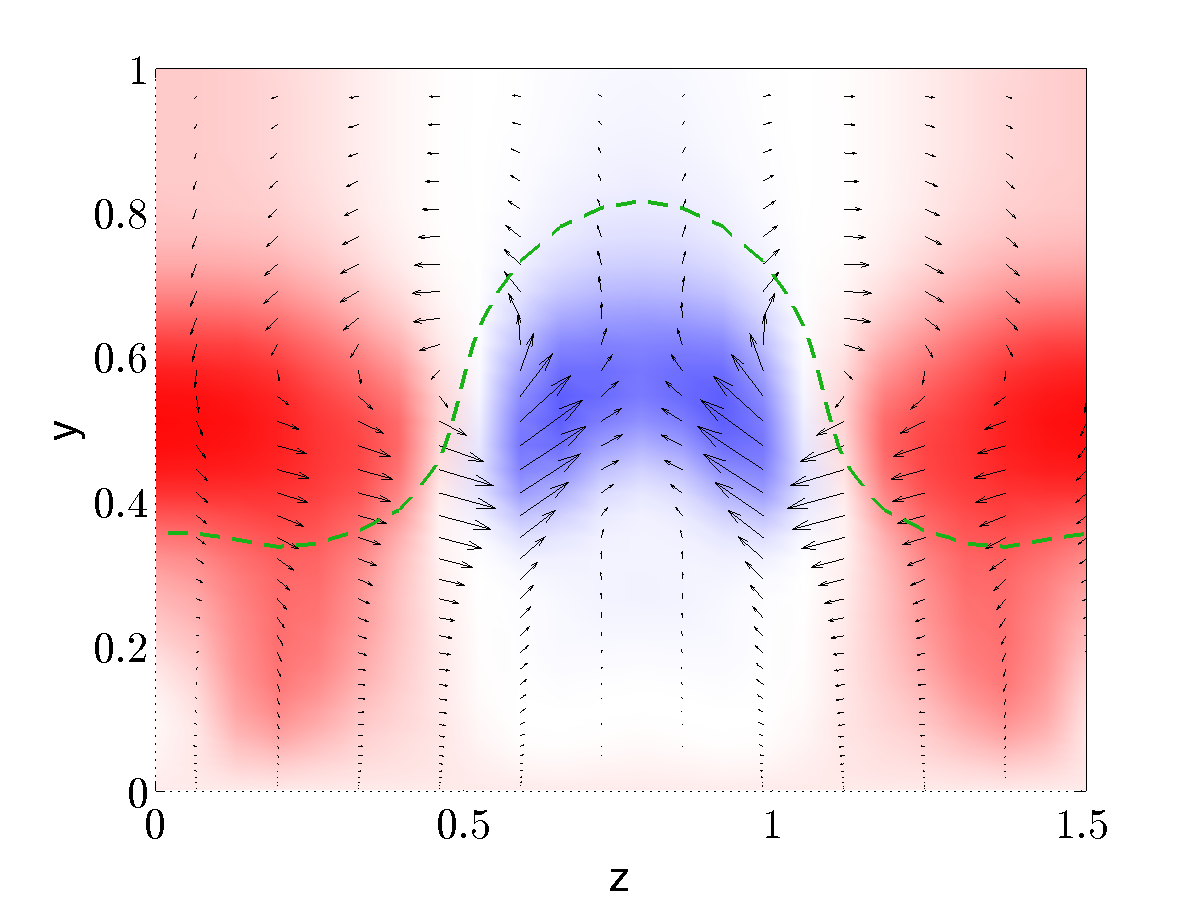}
    \includegraphics[width=0.4\columnwidth]{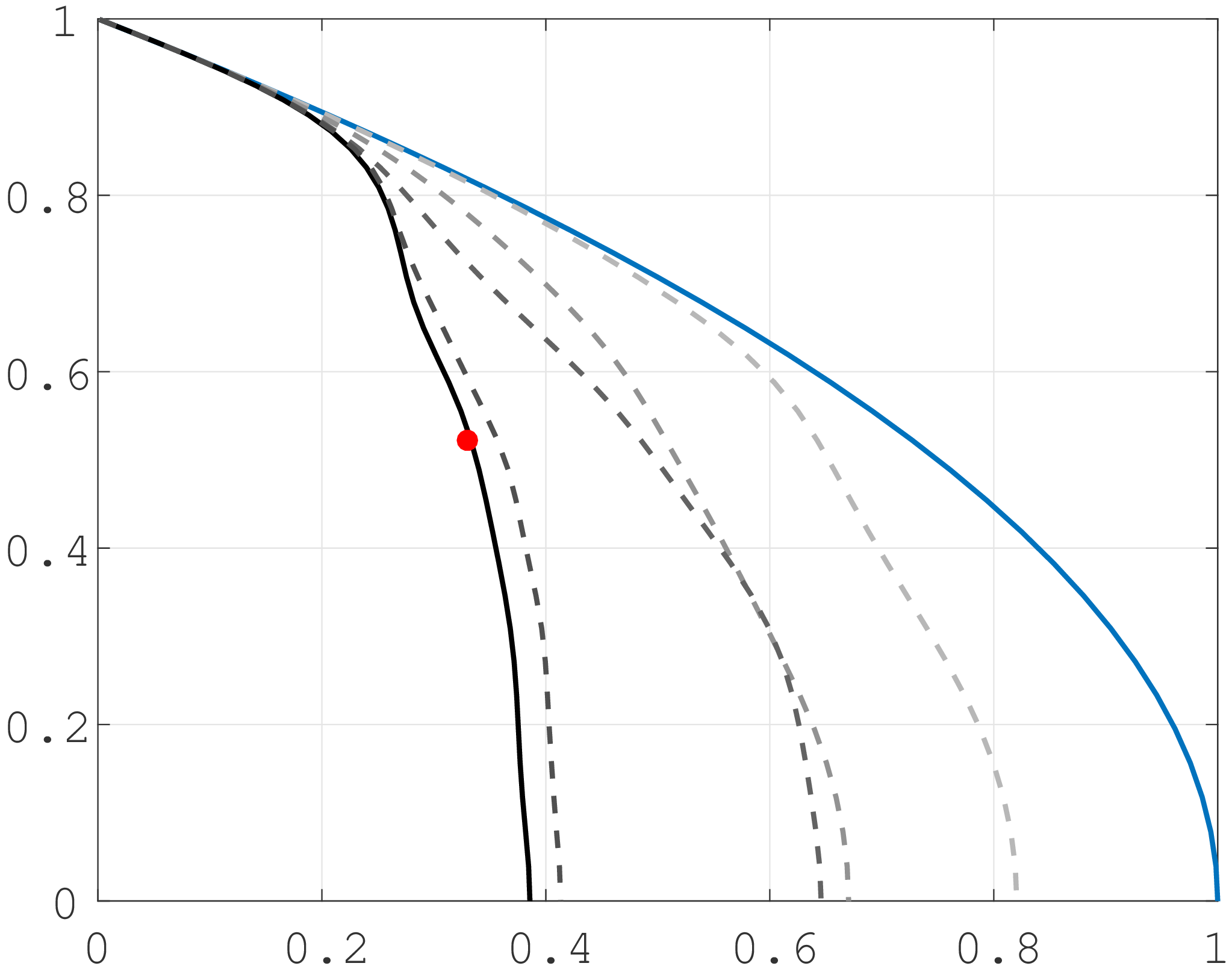}
    \caption{(Color online) P4 upper branch solution in a channel. From left to right: actual solution; projection onto five response modes pairs per Fourier mode (containing 77\% of the fluctuation energy), projection onto one response mode pair per Fourier mode (containing 59\% of the fluctuation energy); mean velocity profile.}
    \label{channel-proj-P4U}
\end{figure*}

The efficiency of all the projections of the pipe and channel solutions are shown in Figure~\ref{proj-E-all}, along with details of the solutions.
The response modes for the channel come in pairs that have odd and even symmetry about the centreline, so the projections are listed using pairs of modes, in accordance with this.
Plots representative of cases of interest for the solution velocity fields projected onto the left singular vectors of the model are shown in figures~\ref{pipe-proj-6507-1000} to~\ref{channel-proj-P4U}.

From the projections, we find that all the lower-branch pipe solutions, and one of the upper branch pipe solutions, are captured very well by one response mode per Fourier mode. In this sense, the model predicts the wall normal form of the velocity fluctuations. We also find, in particular with the P4 channel solution, that the fluctuation energy is typically concentrated around the instantaneous critical layer, where the phase velocity equals the instantaneous velocity. This mechanism is known to be well captured by the model via the average critical layer \cite{McKeonSharma2010, SharmaMcKeon2013}. The extent to which the instantaneous critical layer deviates from the average critical layer depends on the solution in question.

The other two upper branch pipe solutions require more modes to achieve fidelity. The N4U upper branch solution is the most poorly represented solution investigated, with only 80\% of the fluctuation energy captured by the first ten response modes per Fourier mode. We do not know why it is relatively so poorly captured, but recent projections of the turbulent attractor show that 
this invariant solution
is strongly repelling \cite{WillisShort2015}. It is also noticeable that its mean velocity profile looks entirely unlike that of either the turbulent or laminar flow.
The P3U solution is also relatively poorly captured. Examination of this solution shows that it has a relatively fine structure, with energy at many Fourier modes. Similarly, we see that the solutions with more activity at small scales, and so higher dissipation, require mode response modes, since the leading modes tend to be smoother.

Calculations of the internal dissipation of the projected solutions show that the true dissipation of the solutions is quite well captured for most solutions (Figure \ref{proj-eps-all}).
The best-represented solutions (in the sense of dissipation) capture almost all of the dissipation in the first mode. These solutions are close to laminar. The least well captured solutions are typically very energetic at higher wavenumbers and so are more dissipative. These solutions are also the least well captured energetically. We suspect that these solutions require higher dissipation to stabilise them dynamically, meaning more energy is scattered into higher and more dissipative spatial modes. The lower-order projections are smoother and fail to capture this finer structure.
Further calculations of the skin friction for the pipe solution projections (relative to laminar, not shown) show similar results.

\begin{figure*}[]
\centering
\begin{center}
    \begin{tabular}{cc}
\\[0cm]
   \includegraphics[width=0.9\columnwidth]{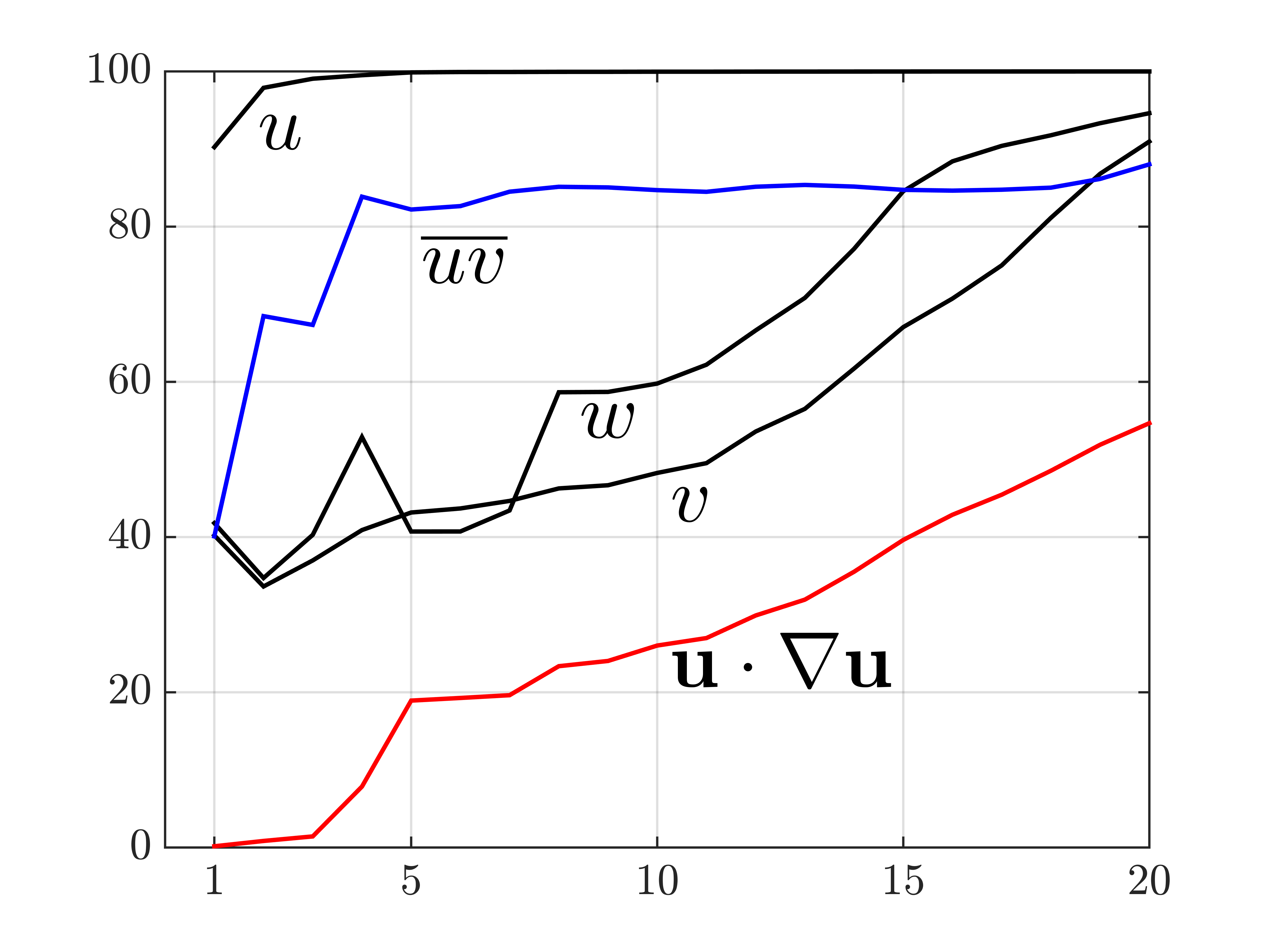}
&
\includegraphics[width=0.9\columnwidth]{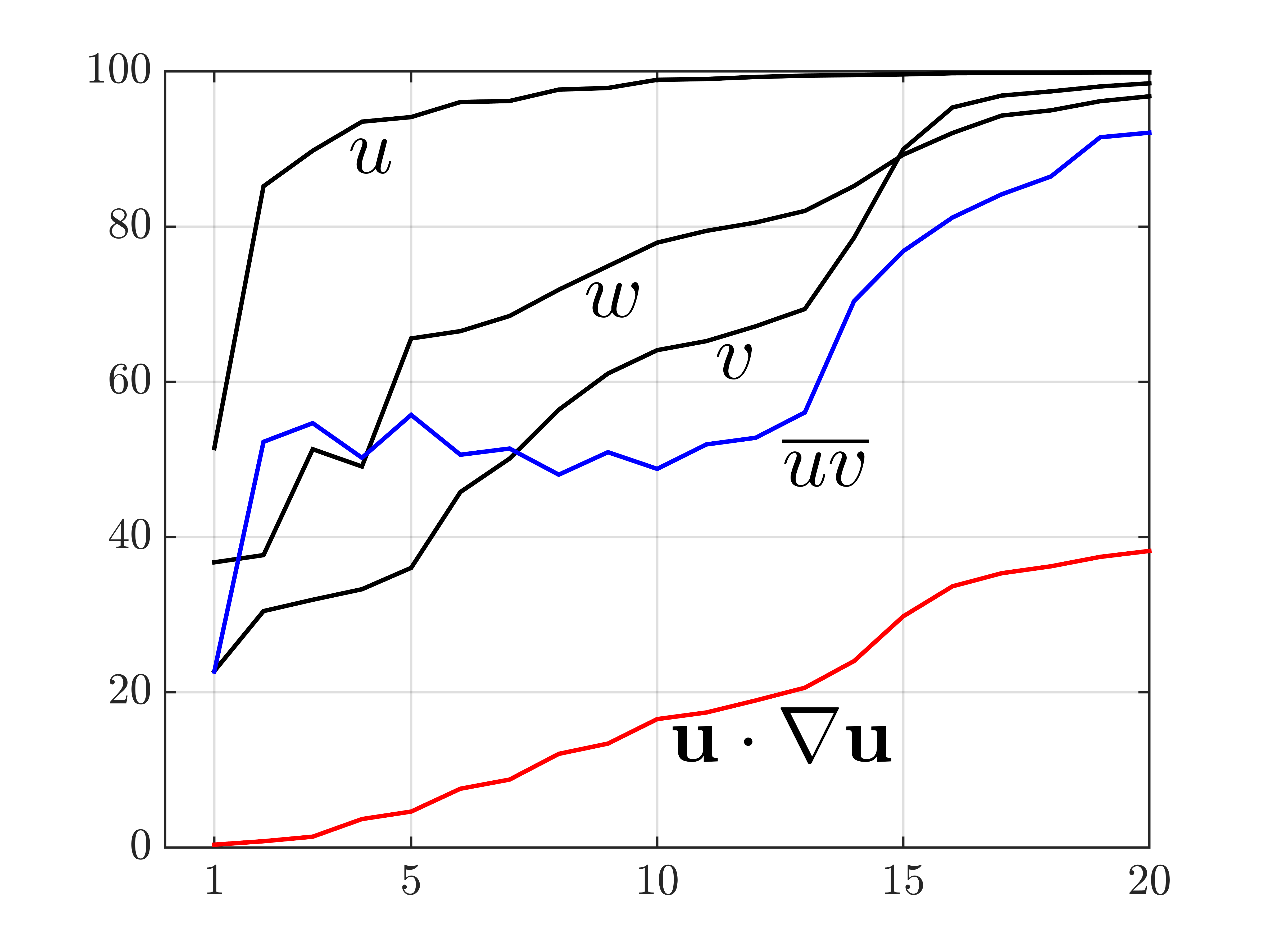}
\\[-2.3cm]
\hspace{-5cm}
\begin{turn}{90}
$$
\end{turn}
&
\hspace{-5cm}
\begin{turn}{90}
$$
\end{turn}
\\[1.6cm]
included response mode pairs
&
included response mode pairs
\\[0.1cm]
$(a)$
&
$(b)$
\\[-0.1cm]
\end{tabular}
\end{center}
\caption{The percentage captured for velocity components ($u, v, w$), for nonlinear forcing ($\bu \cdot \nabla \bu$), and for the $\overline{uv}$ Reynolds stress using increasing numbers of pairs of leading response modes in the P4L (a) and P4U (b) solutions.}
\label{fig.capture-percent}
\end{figure*}

In the following, we pay special attention to the upper and lower branches of the P4 solutions; similar results are observed in the pipe flow. As shown in~\cite{pargra15}, the mean velocities of P4U and P4L are respectively similar to von K{\'a}rm{\'a}n's turbulent profile and Virk's profile for maximum drag reduction by additive polymers. In addition, the normal or `active' turbulent trajectory lies within a region close to the P4U solution while the turbulent trajectory occasionally escapes from this region and approaches the P4L solution. 

Figure~\ref{fig.capture-percent} shows the captured velocity components ($u, v, w$), produced nonlinear forcing ($\bu \cdot \nabla \bu$), and the produced $\overline{uv}$ Reynolds stress using up to 20 most amplified response mode pairs in both P4 solutions. In both solutions, the first few response modes tend to capture the streamwise velocity more than the wall-normal and spanwise velocities, because the streamwise velocity dominates the velocity fluctuations. In addition, the same order of projection captures a relatively smaller portion of $u$ in the high-drag solution (P4U) compared to the low-drag solution (P4L). This is because the streamwise velocity constitutes a larger fraction of the fluctuations in P4U than it does in P4L. 

The model reproduces a much smaller portion of the nonlinear forcing. For example, the first 20 most amplified response mode pairs capture less than 60 and 40 percent of $\bu \cdot \nabla \bu$ for the P4L and P4U solutions respectively, see figure~\ref{fig.capture-percent}. To some extent, this is expected since most of the nonlinear terms are filtered out by the selectively high-gain linear mechanisms in the NSE. In this sense, we argue that the response modes capture the necessary portion of the nonlinear terms, and therefore, can better our understanding of the scaling role of nonlinearity in the NSE.
The nonlinear forcing produced by the pipe solution projections was also compared. Similar trends were observed for the dissipation of the projections. The mean profiles resulting from various projections are shown in figures \ref{pipe-proj-6507-1000} and \ref{pipe-proj-6502-0001}.

Figure~\ref{fig.capture-percent} also shows how the $\overline{uv}$ Reynolds stress is represented by the most amplified response mode pairs. The captured normal stresses $\overline{uu}$, $\overline{vv}$, and $\overline{ww}$ are not shown since they follow the same trends as $u$, $v$, and $w$ fluctuations. We see that $\overline{uv}$ is captured using fewer response mode pairs in the P4L solution compared to the P4U solution and that fewer response modes are required to sustain the mean velocity in the P4L solution. This is confirmed in figure~\ref{channel-proj-P4L} (right) where the mean velocity induced by the response mode pairs (dashed lines) are compared with the mean velocity from the actual solution.

We have shown that the velocity fluctuations in fully nonlinear exact invariant solutions can be predicted and efficiently represented by a model derived to describe high-$Re$ wall-bounded turbulence. This supports the idea that the same basic mechanisms are present in these invariant solutions as in these turbulent flows. Moreover, it should be noted that the model formulation is equally suited to representing periodic orbits, which it has been argued are likely to be more important in the turbulent regime \cite{KerswellTutty2007, WillisShort2015}.

The methodology studied will greatly help further development of the resolvent formulation of wall turbulence, by providing a simplified environment with a single phase velocity in which to study the nonlinear interactions within the model.
It was shown that, for the more complex solutions, even though the nonlinear terms are generally not fully captured by the forcing modes, the necessary portion of the nonlinear terms from an input-output viewpoint is well captured. This implies that a relatively small portion of the nonlinear terms can pass through the selective filtering action of the high-gain linear mechanisms in the NSE. This observation can be used to distinguish the `active' nonlinear terms from the `inactive' ones and may have significant implications for modeling and control of wall-bounded flows.

Because of the small number of coefficients involved, it is hoped that it will become much cheaper to solve for solutions in coefficient space directly, giving low-order approximate solutions to exact invariant solutions. Thus, we hope that low-order approximate coherent structures synthesised from the model will be used to provide seeds for the search for new invariant solutions that are already close to those solutions. It is hoped this will greatly reduce the computational cost of such searches and the technical difficulty at large flow rates, and is a subject of our future work.

\begin{acknowledgements}
    This work has been supported by the Air Force Office
    of Scientific Research (Flow Interactions and Control Program) under awards
    FA9550-14-1-0042 (AS), FA9550-11-1-0094 and FA9550-15-1-0062 (JSP and MDG) and FA9550-12-1-0469
    (RM and BJM).
\end{acknowledgements}

\bibliography{refs}

\end{document}